# MFM and FORC+ study of switching mechanism in $Co_{25}Pd_{75}$ films

Joseph. B. Abugri[1,2], Billy D. Clark[3], P. B. Visscher[1,4,a)], Jie Gong[5], and Subhadra Gupta[1,6]

[1]*Center for Materials for Information Technology, University of Alabama, Tuscaloosa, Alabama 35487, USA*

[2]*Department of Electrical and Computer Engineering, University of Alabama, Tuscaloosa, Alabama 35487, USA*

[3]*Intel Corporation, California, USA*

[4]*Department of Physics and Astronomy, University of Alabama, Tuscaloosa, Alabama 35487, USA*

[5]*Seagate Technology, Bloomington, MN, USA*

[6]*Department of Metallurgical and Materials Engineering, University of Alabama, Tuscaloosa, Alabama 35487, USA*

[a)] Electronic mail: visscher@ua.edu

Recent research on CoPd alloys with perpendicular magnetic anisotropy (PMA) has suggested that they might be useful as the pinning layer in CoFeB/MgO-based perpendicular magnetic tunnel junctions (pMTJ's) for various spintronic applications such as spin-torque transfer random access memory (STT-RAM). We have previously studied the effect of seed layer and composition on the structure (by XRD, SEM, AFM and TEM) and performance (coercivity) of these CoPd films. These films do not switch coherently, so the coercivity is determined by the details of the switching mechanism, which was not studied in our previous paper. In the present paper, we show that information can be obtained about the switching mechanism from magnetic force microscopy (MFM) together with first order reversal curves (FORC), despite the fact that MFM can only be used at zero field. We find that these films switch by a mechanism of domain nucleation and dendritic growth into a labyrinthine structure, after which the unreversed domains gradually shrink to small dots and then disappear.

## 1. Introduction

Recently, we have focused on CoPd alloys for pinning layers in CoFeB/MgO-based perpendicular magnetic tunnel junctions (pMTJs). A study of composition, thickness and temperature variations[1] led to the optimization of perpendicular magnetic anisotropy (PMA) of the CoPd alloy, with a nominal composition of $Co_{25}$ $Pd_{75}$ having the highest perpendicular anisotropy energy of 7 x $10^5$ ergs/$cm^3$. Although this does not compare well with high PMA $L1_0$ materials such as FePt, it is sufficient to pull 1 nm of CoFeB out of plane, is easy to fabricate by co-sputtering, and has a moderate annealing temperature of 400 $^0$C, which is compatible with CMOS fabrication for spin-transfer torque magnetic random access memory (STT-MRAM).

We have deposited $Co_{25}Pd_{75}$ films onto thermally oxidized silicon substrates by DC magnetron co-sputtering from elemental targets onto four different seed layers as follows: a) Si/SiO2/*MgO (5)* /CoPd (20)/ Ta (5) nm; b) Si/SiO2/*Ta (5)*/CoPd (20) Ta (5) nm; c) Si/SiO2/*Ta (5)/Ru (5)/Ta(5)*/CoPd (20)/Ta (5) nm; and d) Si/SiO2/*Ta (5)/Pd (5)/*CoPd (20)/Ta (5) nm, where the layers in italics are the seed layers. The sputtering conditions are detailed in a previous paper[2].

Of the CoPd films on the four seed layers (MgO, Ta, Ta/Pd, Ta/Ru/Ta) that we have deposited, two (on MgO and Ta), do not have stable perpendicularly magnetized states (they begin to switch already at H=0, i.e. have low remanence), have low coercivity, and therefore would not be useful for a pMTJ device. These materials are also hard to study with MFM, which involves passing a nanoscale magnetic tip along the surface, because the magnetic field of the tip disturbs the local magnetic structure. In the other two films (on Ta/Pd and Ta/Ru/Ta) this disturbance is not severe and we have imaged them in this work.

## 2. Switching mechanism: FORC and MFM

The mechanism by which a thin magnetic film switches when the magnetic field is changed has been studied by various imaging techniques such as MOKE (magneto-optical Kerr effect)[3,4] and Lorentz microscopy[5,6]. These techniques are difficult and time-consuming – it is attractive to use the more accessible technique of magnetic force microscopy (MFM)[7,8], but MFM cannot easily be used in a nonzero magnetic field, and only one of the states along a hysteresis curve (the remanent state, the red dot in Fig. 1) has zero field. For a general point along the curve at a field $H_R$, we must return the field to $H = 0$ in order to scan an MFM image, and this may change the structure of the film. However, we find that in many systems, returning to $H = 0$ does **not** change the structure significantly, and MFM **can** be used to determine the switching mechanism. (We will assume that the field $H_R$ at which we want to determine the structure during switching is negative, as it is in a system with high remanence.) One indicator of whether the structure changes is whether the magnetization changes as we increase the field from $H_R$ to zero. Conveniently, there is a widely-used technique that measures exactly this, called the FORC (First Order Reversal Curve) method[9,10,11]. Two FORCs are shown with the hysteresis loop in Fig. 1. They are measured by decreasing the field from positive saturation to a "reversal field" $H_R$, and then increasing the field, measuring the magnetic moment $M(H, H_R)$ at approximately equally-spaced values of the field $H$. In the schematic Fig. 1, and in many real FORC curves (e.g., Fig. 2), the magnetization does not change very much along the FORC curve from $H_R$ (red dot) to zero (blue dot). Even this small change may be mostly due to reversible rotation of the magnetization rather than irreversible switching or domain wall motion – we could subtract off the reversible part[12] and plot only the (flatter) irreversible FORC curve in Fig. 2, but since the change is small we have not done that.

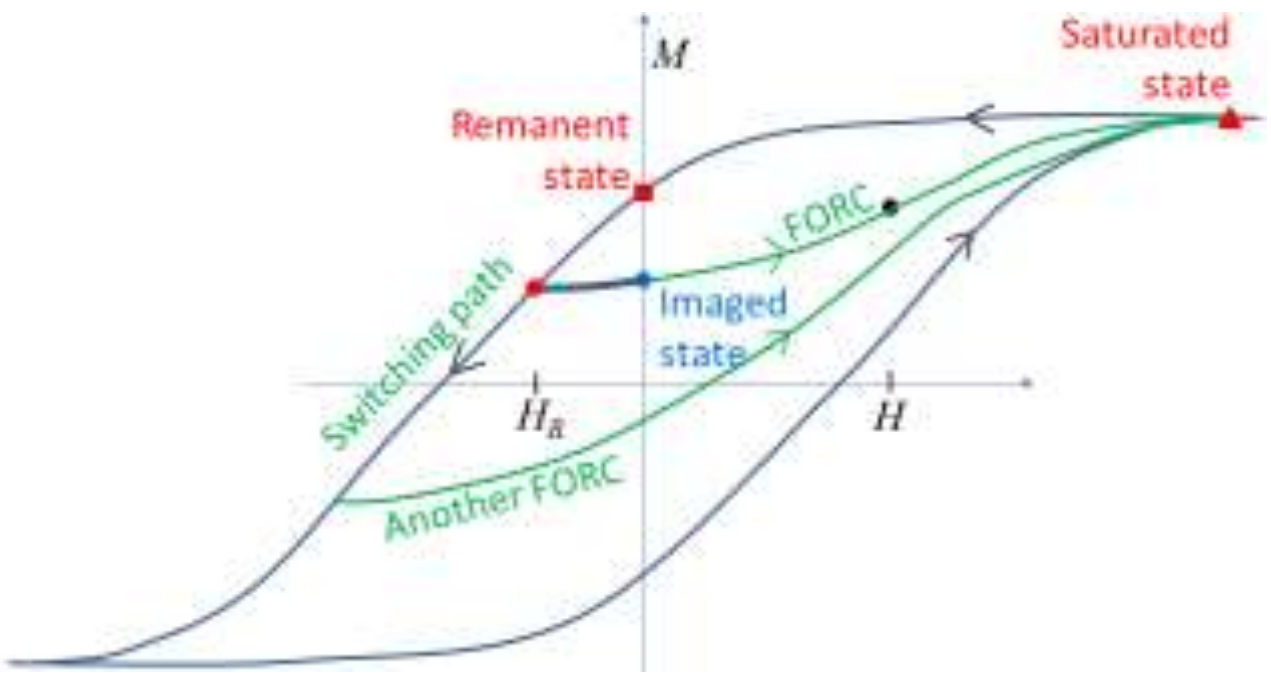

Fig. 1. Schematic hysteresis loop M(H), showing the remanent state (red square) that can be imaged using MFM. Two FORC curves are also shown. On the upper FORC, the red dot at the beginning of the FORC is on the switching path, and when the field is turned off to make an MFM, we move along the FORC to the blue dot.

In an MFM measurement, a vibrating magnetic tip (approximately a vertical point dipole) is swept across the surface of the sample – the changes of vibration phase or amplitude measure approximately[7] $\partial^2 H_z/\partial z^2$. Over a large uniform domain, the stray field **H** vanishes, so the method is primarily sensitive to domain walls. If the sample anisotropy is very small, the coercivity will be small and the passage of the tip can switch grains. Even if the tip field is smaller than the coercivity, it can move a domain wall if it exceeds the domain wall pinning field. This will cause streaking in the image[13]. In our case, of our four samples with different seed layers, the domain wall pinning fields seem to correlate with the coercivity: the highest-coercivity films (with Ta/Pd and Ta/Ru/Ta seeds) have high enough pinning fields to provide a good MFM image. The highest coercivity is obtained with the Ta/Pd seed, whose FORC curves are shown in Fig. 2. It can be seen that the system switches rather abruptly at about -3 kOe, and if we stop at $H_R$, one of the red dots along the switching path (the upper hysteresis loop) and increase the field to zero (blue dot) following the FORC curve, the total moment does not change much.

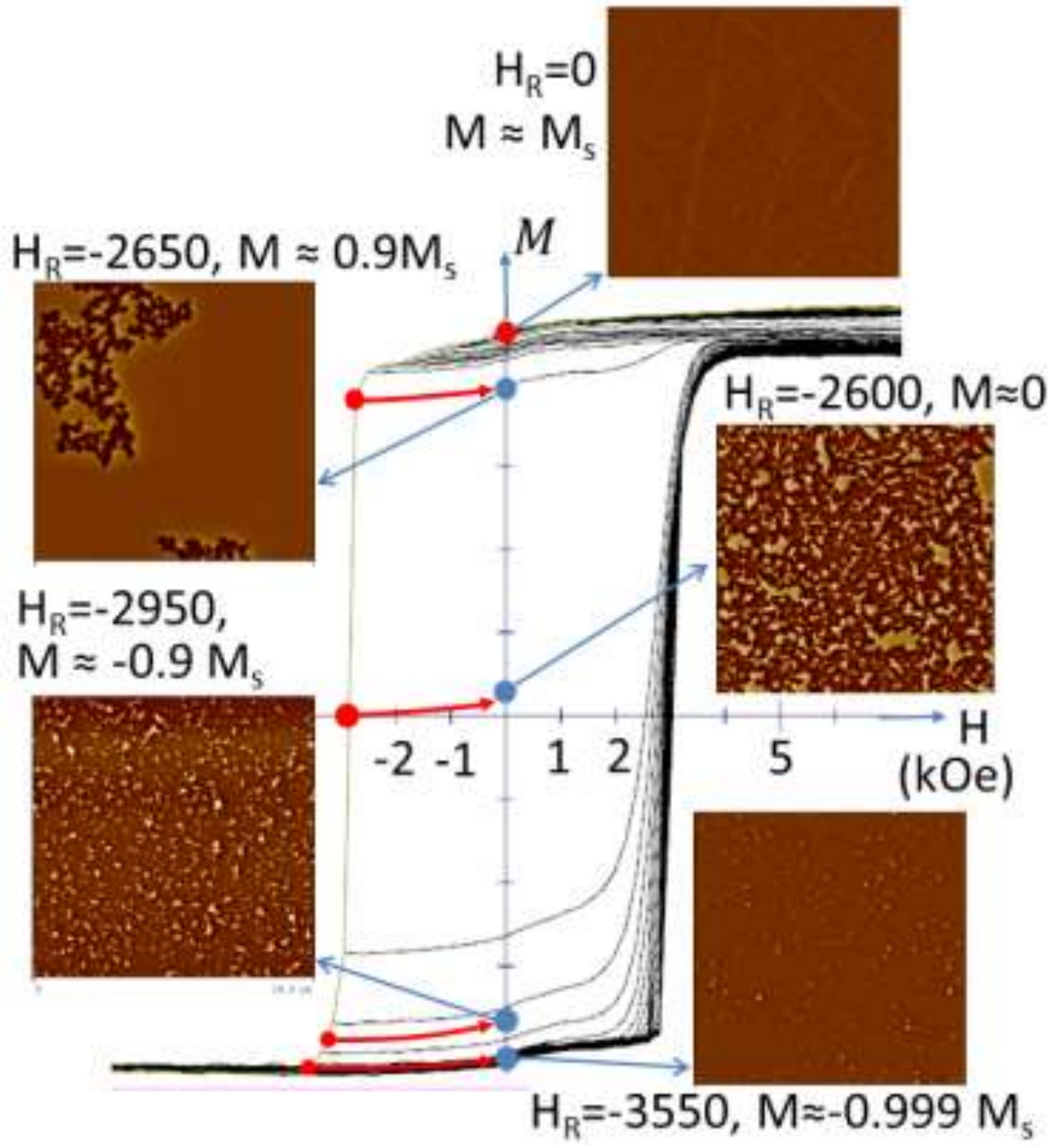

Fig. 2. FORC curves of CoPd alloy film with TaPd seed. The system was brought to each of the 5 red dots along the switching path (at a field $H_R$), then brought to H=0 (blue dot) along the FORC (red curve). The blue arrow from the blue dot points to an inset 10 μm x 10 μm MFM image at the blue dot (H=0). It is difficult to get the moment M in the MFM image to be exactly the same as one of the FORC curves, for example M≈0, but since the FORC curves are very similar, we can sketch (red curve) what the curve would look like at the moment value of the MFM image.

Fig. 2 has insets showing the MFM images at the blue dots. There is no visible change in the saturated state as the field is reduced from ∞ to 0 (upper right image, the remanent state). But the next image (labeled 0.9 $M_s$) shows dendritic growth of reversed domains. Since the magnetization is changing very rapidly with $H_R$ in this region, we have positioned this image in the MH plane by estimating its $M \approx 0.9M_S$ rather than by its $H_R$. It would probably be very difficult to capture the exact field and location of the nucleation event, but the mechanism is clear. The dendritic growth of a reversed region with a spongy texture continues until it fills the area (in the next image, $M \approx 0$). The spongy region then densifies until there are only isolated patches unreversed ($M \approx -0.9M_S$); the last (lower right) image at $H_R = -3550$ Oe has $M \approx -M_S$ but there are still some unreversed spots visible. A last image at $H_R = -4000$ Oe is not shown because it looks completely saturated.

This MFM visualization method does not work for all PMA (perpendicular magnetic anisotropy) films. In particular, the CoPt multilayer film used by Davies *et al*[3] (imaged by MOKE) formed a labyrinthine structure which densifies by domain wall motion as our spongy structure does, but the motion is highly reversible near zero field (a wasp-waisted or "bow-tie" hysteresis loop) so the domain walls move if the field is turned off – MFM can therefore not be used for imaging. The most important reason for this difference may be the thickness – their film was much thicker (50 nm) than ours (20 nm), which increases the magnetostatic interaction fields and makes stripe domains more stable than a uniform saturated state at zero external field. Another notable difference is that rather than the sudden nucleation of reversal in Fig. 2 (so that the actual nuclei cannot be imaged), they observe gradual growth of reversed nuclei as the field is decreased, similar to the patches of unreversed nuclei we see at $H_R = -3550$ Oe.

The traditional use of FORC curves is to compute a "FORC distribution" or "FORC density":

$$\rho(H_R, H) = -\frac{1}{2}\frac{\partial^2 M(H_R, H)}{\partial H\, \partial H_R} \qquad (1)$$

In a system of independent Preisach hysterons[9], this can be interpreted as the density of hysterons that switch down at field $H_R$ and back up at field $H$. In thin films, the grains are far from being independent, but the FORC distribution has still been found to be useful as a "fingerprint" of the system. This is especially true if we can determine the switching mechanism by imaging and identify features in the FORC distribution that are associated with particular mechanisms. This was done by Davies *et al*[3] for a CoPt multilayer film, using MOKE (magneto-optic Kerr effect) for the imaging. They found very fine-scale structure in the FORC distribution which is very difficult to image using existing software, which uses smoothing to suppress noise – the smoothing also suppresses the fine structure. The purpose of the present paper is to (1) show that we can often image switching more easily using MFM and (2) we can visualize very fine structure in the FORC distribution (i.e., avoid the necessity for data averaging) by using a new visualization program (FORC+)[14] which takes advantage of the averaging capability of the human eye by displaying positive and negative density in complementary colors, so noise appears grey from a distance and the positive (orange) and negative (blue) regions are strongly colored and highly visible. The positions of the pixel boundaries in the display are exactly the fields at which measurements were made – no regridding is necessary. The color of each pixel is exactly the discrete partial derivative, involving only data at the four corners of the pixel – no averaging is done.

In Fig. 3 we show a display produced by the FORC+ program – it has the FORC curves at the upper left, and below them the FORC distribution, so the horizontal $H$ axis is the same in both figures. One can then see which features of the FORC curve give rise to the various features of the FORC distribution. Instead of the variables $H$ and $H_R$, it is common to use an equivalent pair of variables, the coercivity and bias field $H_c \equiv \frac{1}{2}(H - H_R)$ and $H_b \equiv \frac{1}{2}(H + H_R)$; these axes are rotated by 45° from the horizontal and vertical axes $H$ and $H_R$, and are shown in Fig. 3.

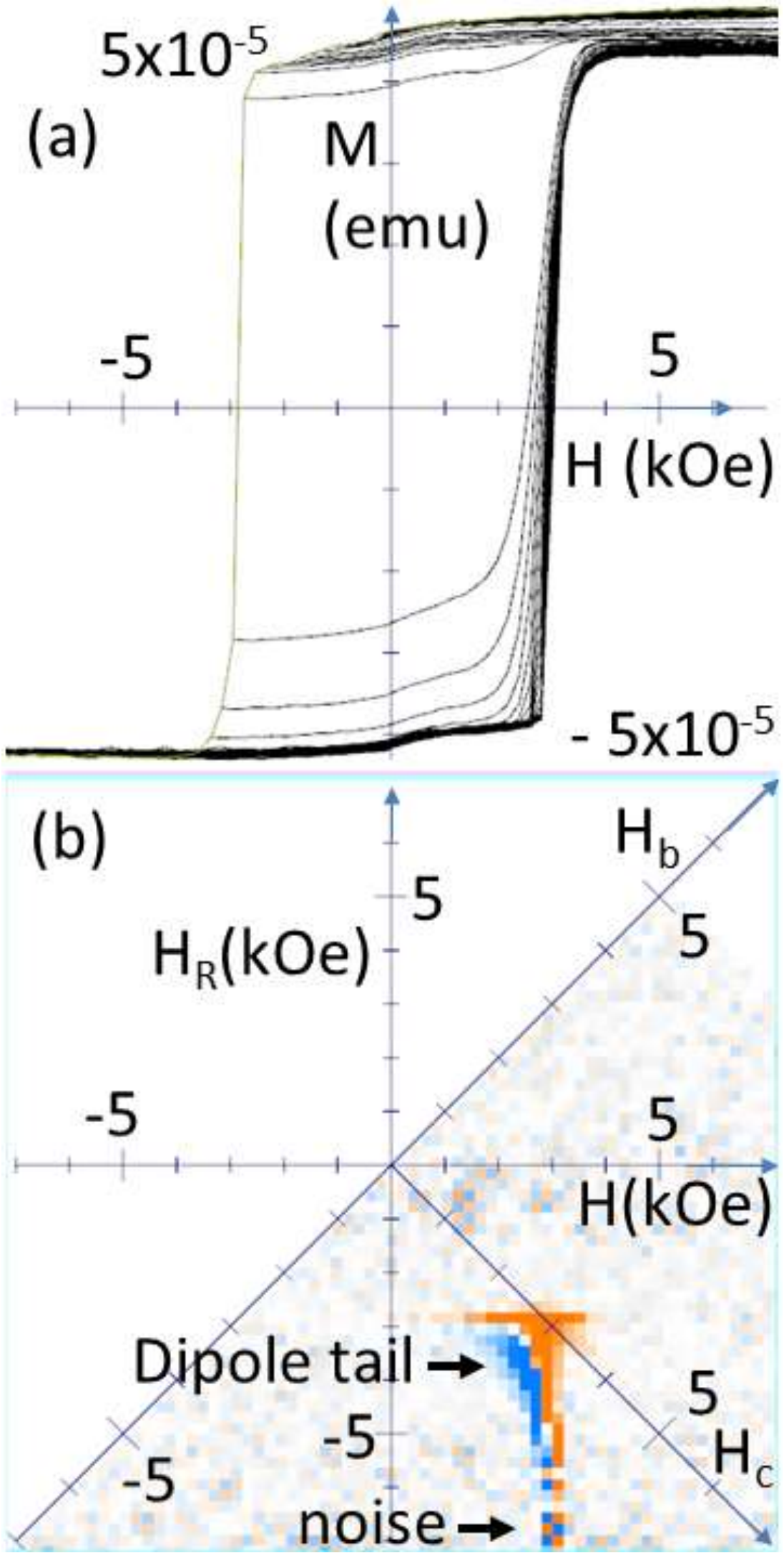

Fig. 3. A display from the FORC+ program for the TaPd seed: (a) FORC curves, (b) FORC density (scale S = $0.4\times10^{-10}$ emu/Oe$^2$, so solid orange means $\rho \geq S$, solid blue is $\rho \leq -S$, half-intensity orange is S/2, etc.) Maximum density is 8S -- the peak is oversaturated in order to show the tail better. Mixed positive and negative pixels below the tail are noise.

The only non-noise structure in Fig. 3(b) is the one we have labeled “dipole tail” because it hangs down along the $H_R$ axis and has a positive (orange) component on the right and a negative (blue) component very close to it on the left. Because they are so close together (one or two pixels, meaning one or two times the field increment $\Delta H$) almost any amount of smoothing will drastically suppress them. For comparison, we show in Fig. 4 the display produced by a widely used FORC display package (VariFORC[15]) that uses averaging over a distance SF (“smoothing factor”) = 5 pixels – note that the dipole tail is hard to see. We could make it more visible by decreasing SF, at the cost of increasing the noise. The small positive and negative features along the upper left ($H_c$=0) boundary are artifacts resulting from not waiting long enough for equilibrium at the beginning of a FORC curve (giving it a “hockey stick” appearance, slightly visible in Fig. 5(a)). They do not appear in our FORC+ distribution (Fig. 3(b)) because FORC+ has an option to replace the first point of each curve by an extrapolation.

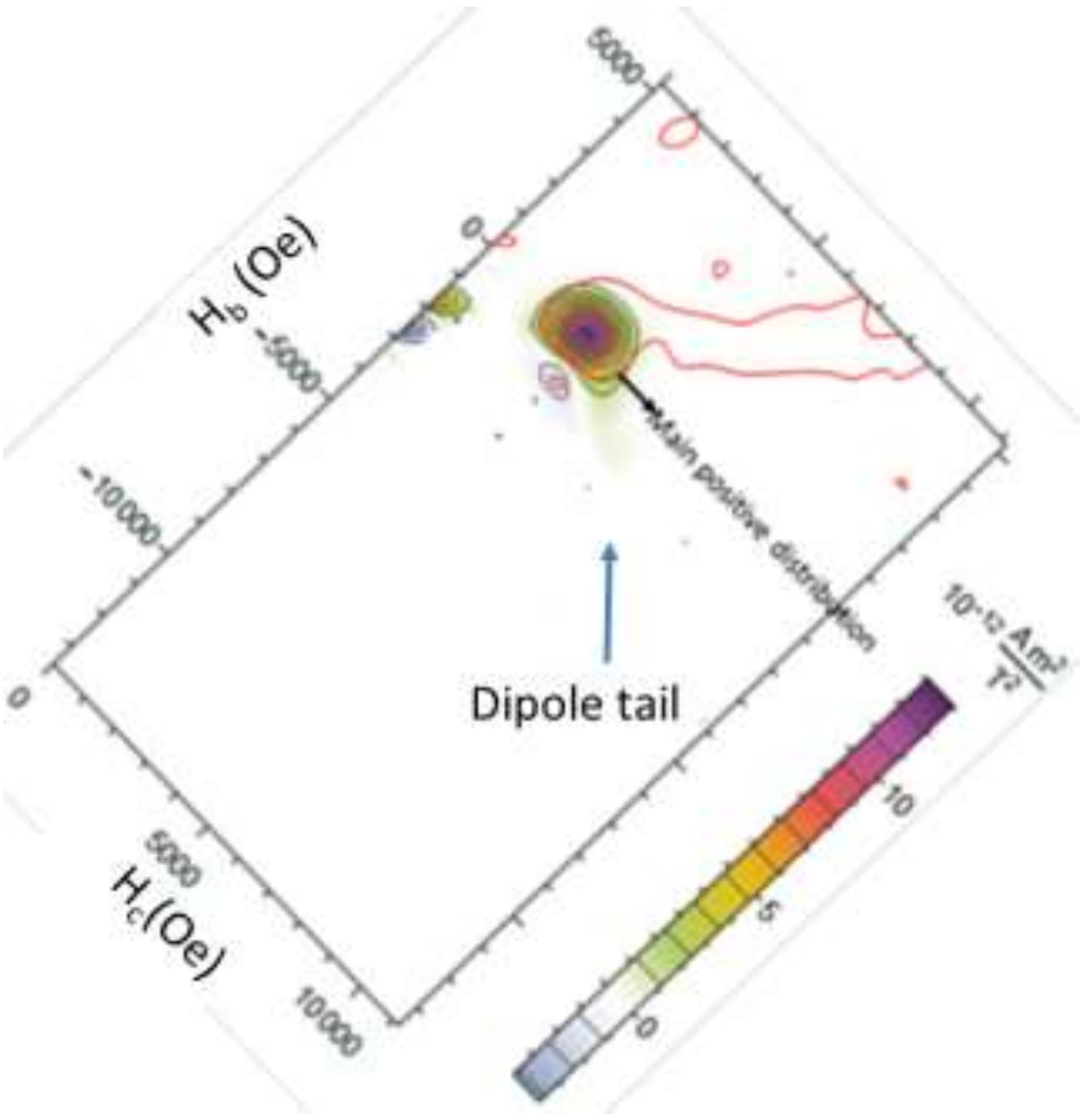

Fig. 4. A smoothed display (produced by VariFORC[15]) for the same system as Fig. 3 (rotated by 45° to match the axes of Fig. 3). The dipole tail is labeled but is very hard to see because of the smoothing.

The dipole tail was observed by Davies *et al*[3], who noted that it has an interesting interpretation. It shows that changes are occurring in the film even after the hysteresis loop seems to be fully saturated. Note that the upper hysteresis curve in Fig. 3(a) is fully saturated at about -3.3 kOe, whereas the tail is clearly visible in the FORC distribution (Fig. 3(b)) at least as low as $H_R = -5$ kOe. [Actually the tail in Fig. 3(b) extends to arbitrarily low $H_R$, but below -5 kOe the positive pixels and negative pixels are mixed, so it is arguably noise. Above -5 kOe the positive density is clearly on the right and it is not noise.] We have observed this also in an MTJ (magnetic tunnel junction) stack[16], and refer to it as “holdout bias” – a few unreversed grains, too small to affect the hysteresis loop (but perhaps visible in the lower right inset in Fig. 2), refuse to switch until about -5 kOe, and change (bias) the field at which the film switches back because they provide nuclei for the back-switching.

The other CoPd film producing a reasonable coercivity and remanence was deposited on the Ta/Ru/Ta seed layer. We show the FORC+ display for this case in Fig. 5. Just as with the Ta/Pd seed, there is a single feature in the FORC distribution, a dipole tail. We can regard this as a “fingerprint” of the dendritic growth + densification switching mechanism, and this conclusion is confirmed by the MFM images for CoPd on Ta/Ru/Ta, which are not shown because they are quite similar to those for CoPd on Ta/Pd (Fig. 2).

The remaining two seed layers produced much smaller remanence and coercivity and thus are probably not useful for applications. We will not present MFM images of these samples because the MFM tip perturbs them and produces streaking (i.e., they have low domain wall pinning fields, as observed above). The FORC+ display for the Ta seed is shown in Fig. 6 – the one for the MgO seed is very similar and is not shown. Note that the FORC structure is very near zero coercivity (the scale is different from Fig. 3) and does not have the clear “dipole tail” form. However, it does have a fingerprint consisting of a positive region just above and to the right of a narrow (one or two pixels) negative stripe – the negative stripe completely disappears upon smoothing, but is clearly visible in the FORC+ display. Note that the density is all at very small field values, so we have zoomed in on a few data points. Of course we could get more resolution by repeating the FORC measurement in this small field range, with a smaller field increment. However, this may not be necessary – FORC+ shows the basic structure even with a very coarse field grid.

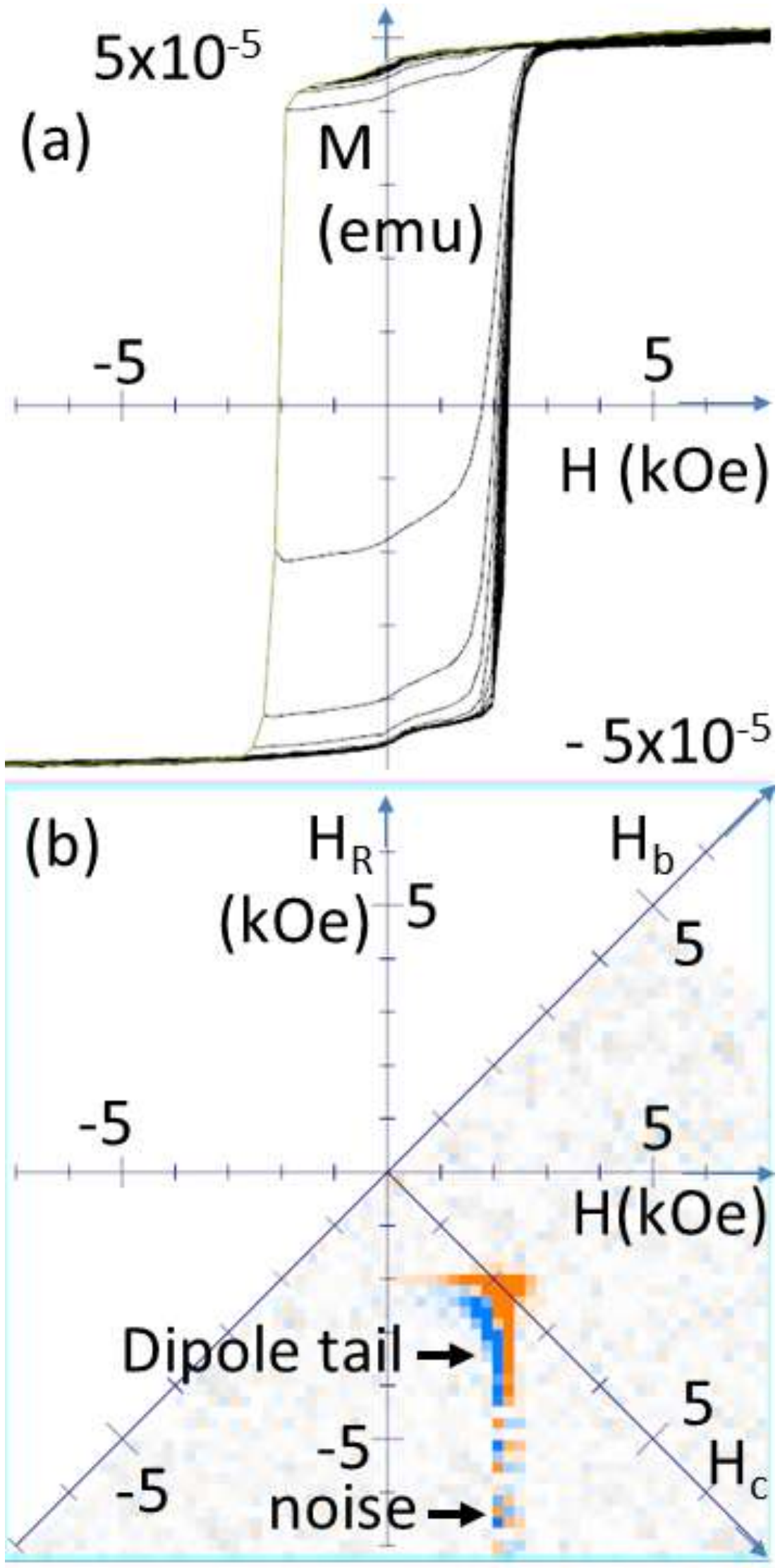

Fig. 5. Display from the FORC+ program, like Fig. 3 but for the Ta/Ru/Ta seed: (a) FORC curves, (b) FORC density (scale S = 0.5x10$^{-10}$ emu/Oe$^2$, maximum density = 8S).

## 3. Conclusion

We have shown from the FORC curves that MFM can be used (even at zero field) to study the intermediate states of some CoPd films during switching and elucidate the switching mechanism. This method seems to work best for thin films (we used 20 nm) for which the demag field is not strong enough to force a stripe structure to form when the external field is turned off. We find that the highest-coercivity $Co_{25}Pd_{75}$ alloy films switch by nucleation and dendritic growth of a region of labyrinthine or spongy domains, within which the unreversed domains shrink, becoming isolated islands, that shrink and disappear as the system approaches negative saturation.

## Acknowledgements

The authors acknowledge Seagate Technology and Eric Singleton for advice and assistance, and the U. of Alabama MINT Center, the Department of Electrical and Computer Engineering, the Department of Physics and Astronomy, and the Department of Metallurgical and Materials Engineering for support. The contributions of the first two authors to this work were of equal importance.

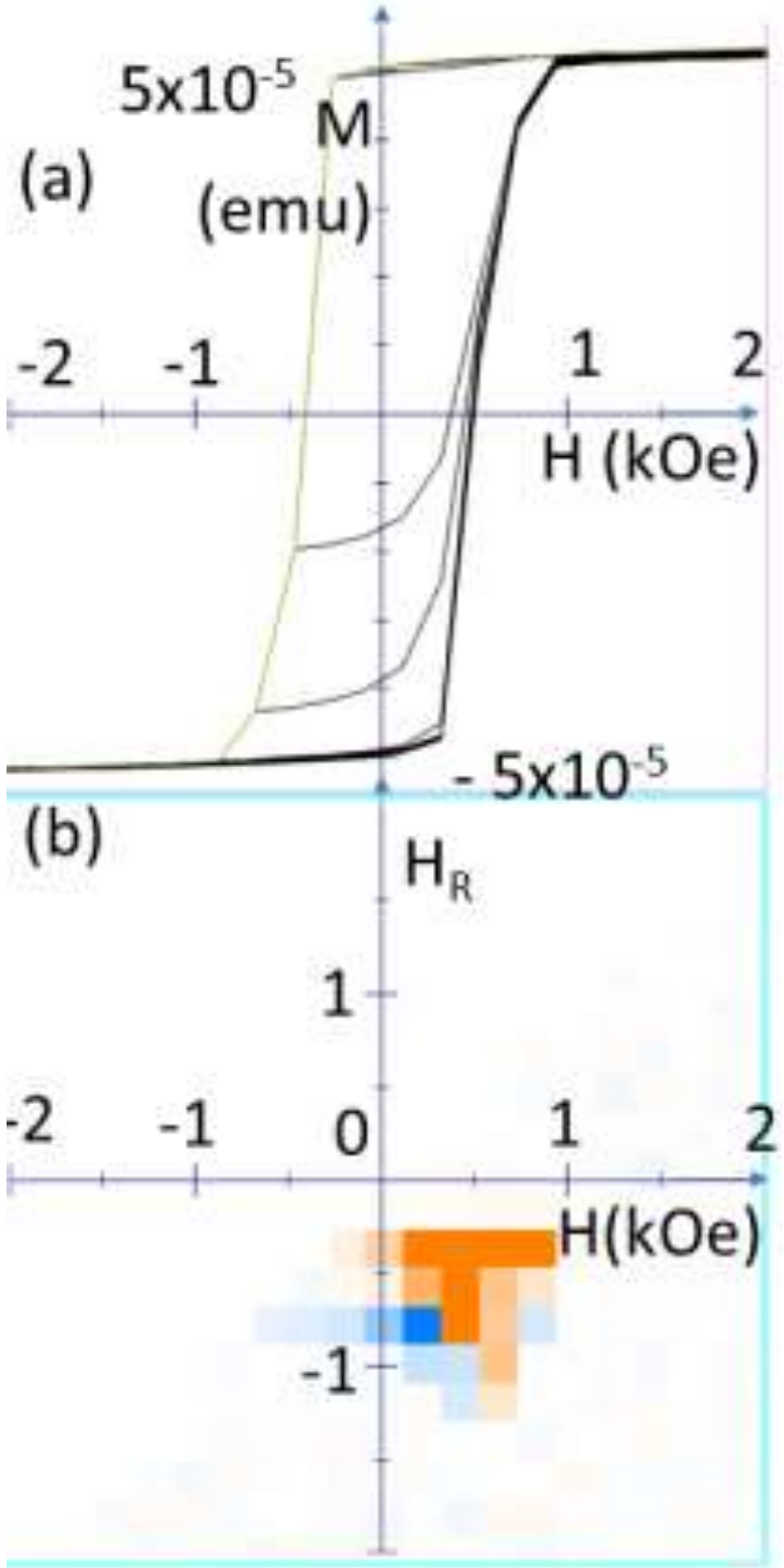

Fig. 6. Display from the FORC+ program, like Fig. 3 but for the Ta seed: (a) FORC curves, (b) FORC diagram (scale S = $1.5 \times 10^{-10}$ emu/Oe$^2$, maximum density 4S).